\documentclass[graybox,vecphys]{svmult}

\usepackage{mathptmx}       
\usepackage{helvet}         
\usepackage{courier}        
\usepackage{type1cm}        
\usepackage{makeidx}         
\usepackage{graphicx}        
\usepackage{multicol}        
\usepackage[bottom]{footmisc}

\usepackage{bm}

\def\br{ \bm{r} }

\def\bk{ \bm{k} }
\def\bK{ \bm{K} }

\def\bq{ \bm{q} }
\def\bA{ \bm{A} }

\def\bgam{ \bm{\gamma} }

\def\im{ \,\mathrm{Im}\, }
\def\re{ \,\mathrm{Re}\, }
\def\sign{ \,\mathrm{sign}\, }
\def\tr{\,\mathrm{tr}\,}
\def\Tr{\,\mathrm{Tr}\,}

\def\bnab{ \bm{\nabla}}
\def\bsigma{ \hat{\bm{\sigma}} }
\def\bD{ \bm{D} }
\def\bH{ \bm{H} }

\makeindex             

\begin{document}

\title*{Effects of impurities in noncentrosymmetric superconductors}

\author{K. V. Samokhin}

\institute{K. V. Samokhin \at Department of Physics, Brock University,
St.Catharines, Ontario L2S 3A1, Canada,\\ \email{kirill.samokhin@brocku.ca}}

\maketitle

\abstract{Effects of disorder on superconducting properties of noncentrosymmetric
compounds are discussed. Elastic impurity scattering, even for scalar impurities, leads to a strongly
anisotropic mixing of the electron states in the bands split by the spin-orbit coupling.
We focus on the calculation of the critical temperature $T_c$, the upper critical field $H_{c2}$, and the spin susceptibility
$\chi_{ij}$. 
It is shown that the impurity effects on the critical temperature are similar to those
in multi-band centrosymmetric superconductors, in particular, Anderson's theorem holds for isotropic singlet pairing.
In contrast, scalar impurities affect the spin susceptibility
in the same way as spin-orbit impurities do in centrosymmetric superconductors.
Another peculiar feature is that in the absence of inversion symmetry scalar disorder
can mix singlet and triplet pairing channels. This leads to significant deviations of the upper critical field
from the predictions of the Werthamer-Helfand-Hohenberg theory in the conventional centrosymmetric case.}

\section{Introduction}
\label{sec: Intro}

Discovery of superconductivity in a heavy-fermion compound CePt$_3$Si (Ref. \cite{Bauer04}) has stimulated considerable interest, both
experimental and theoretical, in the properties of superconductors whose crystal lattice lacks a center of inversion. The list of
noncentrosymmetric superconductors has been steadily growing and
now includes dozens of materials, such as UIr (Ref. \cite{Akazawa04}), CeRhSi$_3$ (Ref. \cite{Kimura05}),
CeIrSi$_3$ (Ref. \cite{Sugitani06}), Y$_2$C$_3$ (Ref.
\cite{Amano04}), Li$_2$(Pd$_{1-x}$,Pt$_x$)$_3$B (Ref. \cite{LiPt-PdB}), and many others. 

A peculiar property of noncentrosymmetric crystals is that the
spin-orbit (SO) coupling of electrons with the crystal lattice qualitatively changes the nature of
the Bloch states, lifting the spin degeneracy of the electron bands almost everywhere in the Brillouin zone. The resulting nondegenerate bands are 
characterized by a complex 
spin texture and a nontrivial wavefunction topology in the momentum space.\cite{Sam09} This has profound consequences for
superconductivity, including unusual nonuniform superconducting phases, both with and without magnetic field,\cite{MinSam94,Agter03,DF03,Sam04,KAS05,MS08}
magnetoelectric effect,\cite{Lev85,Edel89,Yip02,Fuji05,Edel05} and a strongly anisotropic spin susceptibility with a large residual
component at zero temperature.\cite{Edel89,GR01,FAKS04,Sam07} These and other properties are discussed in other chapters of this volume. 

In this chapter we present a theoretical review of the effects of nonmagnetic impurities in superconductors
without inversion symmetry. In Sect. \ref{sec: general}, the disorder-averaged Green's functions in the normal and superconducting 
states are calculated. In Sect. \ref{sec: Tc}, the equations for the superconducting gap functions renormalized by impurities are used to find 
the critical temperature $T_c$. In Sect. \ref{sec: Hc2}, the upper critical field $H_{c2}$ is calculated at arbitrary temperatures. 
In Sect. \ref{sec: susceptibility}, we calculate the spin susceptibility, focusing, in particular, on the effects of impurities on the 
residual susceptibility at $T=0$. Sect. \ref{sec: conclusions} contains a discussion of our results.
Throughtout this chapter we use the units in which $k_B=\hbar=1$.

\section{Impurity scattering in normal and superconducting state}
\label{sec: general}

Let us consider one spin-degenerate band with the dispersion given
by $\epsilon_0(\bk)$, and turn on the SO coupling. The Hamiltonian
of noninteracting electrons in the presence of scalar impurities
can be written in the form $H=H_0+H_{imp}$, where
\begin{equation}
\label{H_0}
    H_0=\sum\limits_{\bk,\alpha\beta}[\epsilon_0(\bk)\delta_{\alpha\beta}+
    \bgam(\bk)\bm{\sigma}_{\alpha\beta}]
    a^\dagger_{\bk\alpha}a_{\bk\beta},
\end{equation}
$\alpha,\beta=\uparrow,\downarrow$ is the spin projection on
the $z$-axis, $\sum_{\bk}$ stands for the summation over the first
Brillouin zone, $\hat{\bm{\sigma}}$ are the Pauli matrices, and the
chemical potential is included in $\epsilon_0(\bk)$.
The ``bare'' band dispersion satisfies $\epsilon_0(-\bk)=\epsilon_0(\bk)$,
$\epsilon_0(g^{-1}\bk)=\epsilon_0(\bk)$, where $g$ is any
operation from the point group $\mathbf{G}$ of the crystal.
The electron-lattice SO coupling is described by the
pseudovector $\bgam(\bk)$, which has the following symmetry
properties: $\bgam(\bk)=-\bgam(-\bk)$,
$g\bgam(g^{-1}\bk)=\bgam(\bk)$. Its momentum dependence crucially depends on $\mathbf{G}$, see Ref. \cite{Sam09}.
For example, in the case of a tetragonal point group $\mathbf{G}=\mathbf{C}_{4v}$, which is realized, e.g., in CePt$_3$Si, CeRhSi$_3$, and
CeIrSi$_3$, the simplest expression for the SO coupling is $\bgam(\bk)=\gamma_0(k_y\hat x-k_x\hat y)$, which is also known as the Rashba model.\cite{Rashba60}
In contrast, in a cubic crystal with $\mathbf{G}=\mathbf{O}$, which describes 
Li$_2$(Pd$_{1-x}$,Pt$_x$)$_3$B, we have $\bgam(\bk)=\gamma_0\bk$.

The impurity scattering is described by the following Hamiltonian:
\begin{equation}
\label{H_imp}
    H_{imp}=\int d^3\br\sum_\alpha
    U(\br)\psi^\dagger_\alpha(\br)\psi_\alpha(\br).
\end{equation}
The impurity  potential $U(\br)$ is a random function with zero
mean and the correlator $\langle
U(\br_1)U(\br_2)\rangle_{imp}=n_{imp}U_0^2\delta(\br_1-\br_2)$, where
$n_{imp}$ is the impurity concentration, and $U_0$ is the strength
of an individual point-like impurity.

The Hamiltonian (\ref{H_0}) can be diagonalized by a unitary
transformation
$a_{\bk\alpha}=\sum_{\lambda}u_{\alpha\lambda}(\bk)c_{\bk\lambda}$,
where $\lambda=\pm$ is the band index (helicity), and
\begin{equation}
\label{u}
    u_{\uparrow\lambda}=\frac{1}{\sqrt{2}}\sqrt{1+\lambda\frac{\gamma_z}{|\bgam|}},\qquad
    u_{\downarrow\lambda}=\lambda\frac{1}{\sqrt{2}}
    \frac{\gamma_x+i\gamma_y}{\sqrt{\gamma_x^2+\gamma_y^2}}\sqrt{1-\lambda\frac{\gamma_z}{|\bgam|}},
\end{equation}
with the following result:
\begin{equation}
\label{H band}
    H=\sum_{\bk}\sum_{\lambda=\pm}\xi_\lambda(\bk)c^\dagger_{\bk\lambda}c_{\bk\lambda}.
\end{equation}
The energy of the fermionic quasiparticles in the $\lambda$th band is given by
$\xi_\lambda(\bk)=\epsilon_0(\bk)+\lambda|\bgam(\bk)|$. This expression is even
in $\bk$ despite the antisymmetry of the SO
coupling, which is a manifestation of the Kramers degeneracy: the
states $|\bk\lambda\rangle$ and $|-\bk,\lambda\rangle$ are related
by time reversal and therefore have the same energy. In real noncentrosymmetric materials, the SO splitting between the helicity
bands is strongly anisotropic. Its magnitude can be characterized
by $E_{SO}=2\max_{\bk}|\bgam(\bk)|$. For instance, in CePt$_3$Si
$E_{SO}$ ranges from 50 to 200 meV (Ref. \cite{SZB04}),
while in Li$_2$Pd$_3$B it is 30 meV, reaching 200 meV in Li$_2$Pt$_3$B (Ref.
\cite{LP05}).

In the band representation the impurity Hamiltonian (\ref{H_imp}) becomes
\begin{equation}
\label{H_imp_band}
    H_{imp}=\frac{1}{{\cal V}}\sum_{\bk\bk'}\sum_{\lambda\lambda'}
    U(\bk-\bk')w_{\lambda\lambda'}(\bk,\bk')
    c^\dagger_{\bk\lambda}c_{\bk'\lambda'},
\end{equation}
where ${\cal V}$ is the system volume,
$U(\bq)$ is the Fourier transform of the impurity potential, and
$\hat w(\bk,\bk')=\hat u^\dagger(\bk)\hat u(\bk')$.
We see that the impurity scattering amplitude in the band representation is momentum-dependent, even for isotropic scalar impurities, and
also acquires both intraband and interband components, the latter causing mixing of the
helicity bands. In the case of a slowly-varying random potential,
keeping only the forward-scattering contribution $U(\bq)\sim\delta_{\bq,0}$, one obtains:
$w_{\lambda\lambda'}(\bk,\bk')=\delta_{\lambda\lambda'}\delta_{\bk,\bk'}$, i.e. the bands are not mixed.

The electron Green's function in the helicity band representation is
introduced in the standard fashion: $G_{\lambda\lambda'}(\bk,\tau;\bk',\tau')=-\langle T_\tau
c_{\bk\lambda}(\tau)c^\dagger_{\bk'\lambda'}(\tau')\rangle$.
In the absence of impurities, we have
$G_{0,\lambda\lambda'}(\bk,\omega_n)=\delta_{\lambda\lambda'}/[i\omega_n-\xi_\lambda(\bk)]$,
where $\omega_n=(2n+1)\pi T$ is the fermionic Matsubara frequency. 

\begin{figure}
    \includegraphics[width=5cm]{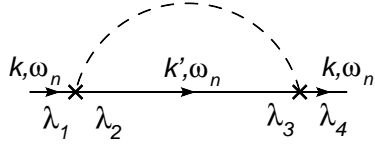}
    \caption{The impurity self-energy in the band representation. The dashed line
    corresponds to $n_{imp}U_0^2$, the vertices include the anisotropy factors $\hat w(\bk,\bk')$,
    and the solid line is the average Green's function of electrons in the normal state. It is shown in the text that the self-energy is nonzero
    only if $\lambda_1=\lambda_4$ and $\lambda_2=\lambda_3$.}
    \label{fig: self-energy}
\end{figure}

We will now show that the impurity-averaged Green's function
remains band-diagonal. The disorder averaging with the Hamiltonian (\ref{H_imp_band}) can
be performed using the standard methods,\cite{AGD} resulting in the Dyson equation of the
form $\hat G^{-1}=\hat G_0^{-1}-\hat\Sigma$, where $\hat G$ is the
average Green's function and $\hat\Sigma$ is the impurity
self-energy, see Fig. \ref{fig: self-energy}. In the Born approximation, taking the thermodynamic limit ${\cal V}\to\infty$, we have
\begin{eqnarray}
\label{Sigma_n_Born}
    \hat\Sigma(\bk,\omega_n)=n_{imp}U_0^2\int\frac{d^3\bk'}{(2\pi)^3}
    \hat w(\bk,\bk')\hat G(\bk',\omega_n)\hat w(\bk',\bk).
\end{eqnarray} 

Seeking solution of the Dyson equation in a band-diagonal form:
$G_{\lambda\lambda'}=G_\lambda\delta_{\lambda\lambda'}$, the
integrand on the right-hand side of Eq. (\ref{Sigma_n_Born}) can be written as follows:
\begin{eqnarray*}
    \hat u(\bk')\hat G(\bk',\omega_n)\hat u^\dagger(\bk')&=&
    \frac{G_+(\bk',\omega_n)+G_-(\bk',\omega_n)}{2}\hat\tau_0\\
    &&+\frac{G_+(\bk',\omega_n)-G_-(\bk',\omega_n)}{2}\hat{\bgam}(\bk')\hat{\bm{\tau}},
\end{eqnarray*}
where $\hat\tau_i$ are the Pauli matrices, and $\hat{\bgam}=\bgam/|\bgam|$. The second
line in this expression vanishes after the momentum integration,
therefore $\hat\Sigma(\bk,\omega_n)=\Sigma(\omega_n)\hat\tau_0$. The
real part of the self-energy renormalizes the chemical potential,
while for the imaginary part we obtain:
$\im\Sigma(\omega_n)=-\Gamma\sign\omega_n$.
Here $\Gamma=\pi n_{imp}U_0^2N_F$ is the elastic scattering rate, with $N_F$ defined as follows: $N_F=(N_++N_-)/2$, where 
$N_\lambda={\cal V}^{-1}\sum_{\bk}\delta[\xi_\lambda(\bk)]$
is the Fermi-level density of states in the $\lambda$th band. Thus we arrive at the following expression for the average 
Green's function of the band electrons:
\begin{equation}
\label{GF_n}
    G_{\lambda\lambda'}(\bk,\omega_n)=
    \frac{\delta_{\lambda\lambda'}}{i\omega_n-\xi_\lambda(\bk)+i\Gamma\sign\omega_n}.
\end{equation}
This derivation is valid under the assumption that the
elastic scattering rate is small compared with the Fermi energy
$\epsilon_F$, which justifies neglecting the diagrams with crossed
impurity lines in the self-energy in Fig. \ref{fig: self-energy}.

\subsection{Impurity averaging in superconducting state}
\label{sec: SC}

In the limit of strong SO coupling, i.e. when the band splitting
exceeds all superconducting energy scales, the Cooper pairing
between the electrons with opposite momenta occurs only if they
are from the same nondegenerate band. The pairing interaction
in the strong SO coupling case is most naturally introduced using
the basis of the exact band states,\cite{GR01,SZB04,SC04}
which already incorporate the effects of the crystal lattice
potential and the SO coupling. The total Hamiltonian including the pairing interaction
is given by $H=H_0+H_{imp}+H_{int}$, where the first two terms are
given by Eqs. (\ref{H_0}) and (\ref{H_imp_band}) respectively, and the
last term has the following form:
\begin{eqnarray}
\label{H int}
    H_{int}=\frac{1}{2{\cal V}}\sum\limits_{\bk\bk'\bq}\sum_{\lambda\lambda'}
    V_{\lambda\lambda'}(\bk,\bk')c^\dagger_{\bk+\bq,\lambda}
    c^\dagger_{-\bk,\lambda}c_{-\bk',\lambda'}c_{\bk'+\bq,\lambda'}.
\end{eqnarray}
Physically, the pairing interaction is mediated by some bosonic excitations, e.g. phonons, and is effective only
at frequencies smaller than a cutoff frequency $\omega_c$, which has to be included in the appropriate Matsubara sums. 
Alternatively, the cutoff can be imposed on the momenta in Eq. (\ref{H int}), as in the original Bardeen-Cooper-Schrieffer (BCS) model. 
The diagonal elements of the pairing potential $V_{\lambda\lambda'}$ describe
the intraband Cooper pairing, while the off-diagonal ones correspond to the pair scattering from one band to the other.

The pairing potential can be represented in the following form:
$V_{\lambda\lambda'}(\bk,\bk')=t_\lambda(\bk)t^*_{\lambda'}(\bk')\tilde V_{\lambda\lambda'}(\bk,\bk')$, see Ref. \cite{SM08}.
Here $t_\lambda(\bk)=-t_\lambda(-\bk)$ are non-trivial phase factors originating in the expression for
the time reversal operation for electrons in the helicity bands:
$K|\bk\lambda\rangle=t_\lambda(\bk)|-\bk,\lambda\rangle$,
\cite{GR01,SC04} while the components of $\tilde V_{\lambda\lambda'}$ are even in both $\bk$ and $\bk'$ and invariant under the point group operations: $\tilde V_{\lambda\lambda'}(g^{-1}\bk,g^{-1}\bk')=\tilde V_{\lambda\lambda'}(\bk,\bk')$. 
The latter can be expressed in terms of the basis functions of the irreducible representations of the point group.\cite{Book} In general, the basis functions are different for each matrix element. 
Neglecting this complication, and also considering only the one-dimensional representation corresponding to the pairing channel with the
maximum critical temperature, one can write
\begin{equation}
\label{pairing potential}
    \tilde V_{\lambda\lambda'}(\bk,\bk')=-V_{\lambda\lambda'}
    \phi_\lambda(\bk)\phi^*_{\lambda'}(\bk'),
\end{equation}
where the coupling constants $V_{\lambda\lambda'}$ form a
symmetric positive-definite $2\times 2$ matrix, and $\phi_\lambda(\bk)$ are even
basis functions. While $\phi_+(\bk)$ and
$\phi_-(\bk)$ have the same symmetry, their momentum dependence
does not have to be the same. The basis functions are assumed to be real and normalized: $\langle|\phi_\lambda(\bk)|^2\rangle_\lambda=1$, where
the angular brackets denote the Fermi-surface averaging in the $\lambda$th band:
$\langle(...)\rangle_\lambda=(1/N_\lambda{\cal V})\sum_{\bk}(...)\delta[\xi_\lambda(\bk)]$.

Treating the pairing interaction (\ref{H int}) in the mean-field approximation, one introduces the superconducting order parameters in the
helicity bands, which have the following form: $\Delta_\lambda(\bk,\bq)=t_\lambda(\bk)\tilde\Delta_\lambda(\bk,\bq)$. 
The superconducting order parameter is
given by a set of complex gap functions, one for each band, which
are coupled due to the interband scattering of the Cooper pairs
and other mechanisms, e.g. impurity scattering. Thus the overall
structure of the theory resembles that of multi-band superconductors.\cite{SMW59,Moskal59} 
If the pairing corresponds 
to a one-dimensional representation, see Eq. (\ref{pairing potential}), then we have $\tilde\Delta_\lambda(\bk,\bq)=\eta_\lambda(\bq)\phi_\lambda(\bk)$.

Important particular case is a BCS-like model in which the pairing interaction is local in real space:
\begin{equation}
\label{H-BCS}
    H_{int}=-V\int d^3\br\,\psi_\uparrow^\dagger(\br)\psi_\downarrow^\dagger(\br)\psi_\downarrow(\br)\psi_\uparrow(\br),
\end{equation}
where $V>0$ is the coupling constant. One can show\cite{SM08} that in this model there is no interband pairing for any strength of the SO coupling,
the order parameter has only one component $\eta$, 
the gap symmetry corresponds to the unity representation with $\phi_\lambda(\bk)=1$, and all coupling
constants in Eq. (\ref{pairing potential}) take the same value: $V_{\lambda\lambda'}=V/2$.

Let us calculate the impurity-averaged Green's functions in the superconducting state. To make notations compact, the normal and anomalous 
Green's functions\cite{AGD} can be combined into a $4\times 4$ matrix
${\cal G}(\bk_1,\bk_2;\tau)=-\langle T_\tau C_{\bk_1}(\tau)C^\dagger_{\bk_2}(0)\rangle$, where
$C_{\bk}=(c_{\bk\lambda},c^\dagger_{-\bk,\lambda})^T$ are four-component Nambu operators.
Averaging with respect to the impurity positions restores translational invariance: $\langle{\cal
G}(\bk_1,\bk_2;\omega_n)\rangle_{imp}=\delta_{\bk_1,\bk_2}{\cal
G}(\bk,\omega_n)$, where
\begin{equation}
\label{matrix G}
    {\cal G}(\bk,\omega_n)=\left(\begin{array}{cc}
        \hat G(\bk,\omega_n) & -\hat F(\bk,\omega_n) \\
        -\hat F^\dagger(\bk,\omega_n) & -\hat G^T(-\bk,-\omega_n) \\
    \end{array}\right),
\end{equation}
and the hats denote $2\times 2$ matrices in the band space.
The average matrix Green's function satisfies the Gor'kov
equations: $({\cal G}_0^{-1}-\Sigma_{imp}){\cal G}=1$, where
\begin{equation}
\label{matrix G0}
    {\cal G}_0^{-1}(\bk,\omega_n)=\left(\begin{array}{cc}
    i\omega_n-\hat\xi(\bk) & -\hat\Delta(\bk) \\
    -\hat\Delta^\dagger(\bk) & i\omega_n+\hat\xi(\bk)\\
    \end{array}\right),
\end{equation}
and the impurity self-energy in the self-consistent Born approximation is
\begin{eqnarray}
\label{Sigma imp}
    \Sigma_{imp}(\bk,\omega_n)=n_{imp}U_0^2\int\frac{d^3\bk'}{(2\pi)^3}W(\bk,\bk')
    {\cal G}(\bk',\omega_n)W(\bk',\bk),
\end{eqnarray}
which is the Nambu-matrix generalization of Eq. (\ref{Sigma_n_Born}). The $4\times 4$ matrix $W$ is defined as follows:
$W(\bk,\bk')=\mathrm{diag}\,[\hat w(\bk,\bk'),-\hat w^T(-\bk',-\bk)]$.
It is straightforward to show that $[\hat w^T(-\bk',-\bk)]_{\lambda\lambda'}=t^*_\lambda(\bk)t_{\lambda'}(\bk')
w_{\lambda\lambda'}(\bk,\bk')$. We assume the disorder to be sufficiently weak, so that
it is legitimate to use the Born approximation. Although there are some interesting qualitative effects in the opposite limit of strong 
disorder, such as the impurity resonance states,\cite{LE08} these are beyond the scope of our study.  

In the absence of impurities, the Green's functions have the following form:
$G_{0,\lambda\lambda'}(\bk,\omega_n)=\delta_{\lambda\lambda'}G_{0,\lambda}(\bk,\omega_n)$, and 
$F_{0,\lambda\lambda'}(\bk,\omega_n)=\delta_{\lambda\lambda'}t_\lambda(\bk)\tilde F_{0,\lambda}(\bk,\omega_n)$, where
\begin{equation}
\label{GFs clean}
    G_{0,\lambda}=-\frac{i\omega_n+\xi_\lambda}{\omega_n^2
    +\xi_\lambda^2+|\tilde\Delta_\lambda|^2},\quad
    \tilde F_{0,\lambda}=\frac{\tilde\Delta_\lambda}{\omega_n^2
    +\xi_\lambda^2+|\tilde\Delta_\lambda|^2}.
\end{equation}
In the presence of impurities, we seek solution of the Gor'kov equations in a band-diagonal form and
require, for consistency, that the Nambu matrix components of the self-energy are also band-diagonal. Then,
\begin{eqnarray}
    &&\left(\begin{array}{cc}
    \Sigma_{\lambda\lambda'}^{11}(\bk,\omega_n) & \Sigma_{\lambda\lambda'}^{12}(\bk,\omega_n) \\
    \Sigma_{\lambda\lambda'}^{21}(\bk,\omega_n) & \Sigma_{\lambda\lambda'}^{22}(\bk,\omega_n) \\
    \end{array}\right)=\delta_{\lambda\lambda'}\left(\begin{array}{cc}
    \Sigma_1(\omega_n) & t_\lambda(\bk)\Sigma_2(\omega_n) \\
    t_\lambda^*(\bk)\Sigma_2^*(\omega_n) & -\Sigma_1(-\omega_n) \\
    \end{array}\right),
\end{eqnarray}
where $\Sigma_1$ and $\Sigma_2$ satisfy the equations
\begin{equation}
\label{Sigma-12}
    \begin{array}{l}
    \displaystyle\Sigma_1(\omega_n)=\frac{1}{2}n_{imp}U_0^2\sum_\lambda\int\frac{d^3\bk}{(2\pi)^3}
    G_\lambda(\bk,\omega_n), \\
    \displaystyle\Sigma_2(\omega_n)=\frac{1}{2}n_{imp}U_0^2\sum_\lambda\int\frac{d^3\bk}{(2\pi)^3}
    {\tilde F}_\lambda(\bk,\omega_n).
    \end{array}
\end{equation}
Absorbing the real part of $\Sigma_1$ into the chemical potential,
we have $\Sigma_1(\omega_n)=i\tilde\Sigma_1(\omega_n)$, where
$\tilde\Sigma_1$ is odd in $\omega_n$.

Solving the Gor'kov equations we obtain the following expressions for the disorder-averaged Green's functions:
\begin{equation}
\label{GFs average}
    \begin{array}{l}
    \displaystyle G_\lambda(\bk,\omega_n)=-\frac{i\tilde\omega_n+\xi_\lambda(\bk)}{\tilde\omega_n^2
    +\xi_\lambda^2(\bk)+|D_\lambda(\bk,\omega_n)|^2},\\ \\
    \displaystyle \tilde F_\lambda(\bk,\omega_n)=\frac{D_\lambda(\bk,\omega_n)}{\tilde\omega_n^2
    +\xi_\lambda^2(\bk)+|D_\lambda(\bk,\omega_n)|^2},
    \end{array}
\end{equation}
where $\tilde\omega_n=\omega_n-\tilde\Sigma_1(\omega_n)$ and
$D_\lambda(\bk,\omega_n)=\tilde\Delta_\lambda(\bk)+\Sigma_2(\omega_n)$.
Substituting these into Eqs. (\ref{Sigma-12}), we
arrive at the self-consistency equations for the Matsubara frequency and the gap functions renormalized by impurities:
\begin{eqnarray}
\label{tilde omega eq}
    &&\tilde\omega_n=\omega_n+\frac{\Gamma}{2}\sum_\lambda\rho_\lambda
    \left\langle\frac{\tilde\omega_n}{\sqrt{\tilde\omega_n^2+|D_\lambda(\bk,\omega_n)|^2}}
    \right\rangle_\lambda,\\
\label{D eq uniform}
    &&D_\lambda(\bk,\omega_n)=\eta_\lambda\phi_\lambda(\bk)+\frac{\Gamma}{2}\sum_{\lambda'}\rho_{\lambda'}
    \left\langle\frac{D_{\lambda'}(\bk',\omega_n)}{\sqrt{\tilde\omega_n^2
    +|D_{\lambda'}(\bk',\omega_n)|^2}}\right\rangle_{\lambda'},
\end{eqnarray}
where
\begin{equation}
\label{rhos}
    \rho_\pm=\frac{N_\pm}{N_F}=1\pm\delta
\end{equation}
are the fractional densities of states in the helicity bands. The parameter $\delta=(N_+-N_-)/(N_++N_-)$ characterizes the strength of the 
SO coupling.

\section{Gap equations and the critical temperature}
\label{sec: Tc}

The Gor'kov equations must be supplemented by self-consistency equations for the order parameter components, which 
have the form usual for two-band superconductors. In particular, for a uniform order parameter, $\eta_\lambda(\bq)=\eta_\lambda\delta(\bq)$, we have
$\sum_{\lambda'}V^{-1}_{\lambda\lambda'}\eta_{\lambda'}=T\sum_n\int_{\bk}\tilde F_\lambda(\bk,\omega_n)\phi_\lambda(\bk)$ (recall that the basis functions
are assumed to be real). Using Eq. (\ref{GFs average}), we obtain:
\begin{equation}
\label{gap-eq gen}
    \sum_{\lambda'}V^{-1}_{\lambda\lambda'}\eta_{\lambda'}=\pi\rho_\lambda N_F T\sum_n{}'
    \left\langle\frac{D_\lambda(\bk,\omega_n)\phi_\lambda(\bk)}{\sqrt{\tilde\omega_n^2+|D_\lambda(\bk,\omega_n)|^2}}\right\rangle_\lambda.
\end{equation}
These equations are called the gap equations and, together with Eqs. (\ref{tilde omega eq}) and (\ref{D eq uniform}), completely determine the properties
of disordered noncentrosymmetric superconductors in the uniform state. 
The prime in the Matsubara sum means that the summation is limited to $\omega_n\leq\omega_c$, where $\omega_c$ is the BCS frequency cutoff. 

The superconducting critical temperature can be found from Eq. (\ref{gap-eq gen}) after linearization with respect to the
order parameter components. It follows from Eq. (\ref{tilde omega eq}) that $\tilde\omega_n=\omega_n+\Gamma\sign\omega_n$ near $T_c$, and we obtain from 
Eq. (\ref{D eq uniform}) that $\Sigma_2(\omega_n)=(\Gamma/2|\omega_n|)\sum_\lambda\rho_\lambda\langle\phi_\lambda\rangle\eta_\lambda$ 
(here and below we omit, for brevity, the arguments of the basis functions and the subscripts $\lambda$ in the Fermi-surface averages). 
Therefore the linearized gap equations take the form $\sum_{\lambda'}a_{\lambda\lambda'}\eta_{\lambda'}=0$, where
\begin{eqnarray}
\label{a-uniform-gen}
    a_{\lambda\lambda'}=\frac{1}{N_F}V^{-1}_{\lambda\lambda'}
    -\rho_\lambda\delta_{\lambda\lambda'}S_{01}-\frac{\Gamma}{2}\rho_{\lambda}\rho_{\lambda'}
    \langle\phi_{\lambda}\rangle\langle\phi_{\lambda'}\rangle S_{11},
\end{eqnarray}
with $S_{kl}=\pi T\sum_n|\omega_n|^{-k}(|\omega_n|+|\Gamma)^{-l}$. The Matsubara sums here can be easily calculated:
$$
    S_{01}=\pi T\sum_n{}'\frac{1}{\omega_n+\Gamma}=\ln\frac{2e^{C}\omega_c}{\pi T}
	-\mathcal{F}\left(\frac{T}{\Gamma}\right),
$$
where $C\simeq 0.577$ is Euler's constant,
\begin{equation}
\label{mathbb F}
    \mathcal{F}(x)=\Psi\left(\frac{1}{2}+\frac{1}{2\pi x}\right)-\Psi\left(\frac{1}{2}\right),
\end{equation}
and $\Psi(x)$ is the digamma function. Note that the expression (\ref{mathbb F}) for the impurity correction to
$S_{01}$ is valid if $\Gamma\ll\omega_c$, when it is legitimate to extend the summation in $2\pi
T\sum_n[1/(\omega_n+\Gamma)-1/\omega_n]$ to infinity and express
the result in terms of the digamma functions, see Ref. \cite{AM-review}. Similarly, we obtain: $S_{11}=\mathcal{F}(T/\Gamma)/\Gamma$. 
It is convenient to introduce the following notation for dimensionless coupling constants:
\begin{equation}
\label{hat g def}
    g_{\lambda\lambda'}=N_FV_{\lambda\lambda'}\rho_{\lambda'}=V_{\lambda\lambda'}N_{\lambda'}
\end{equation}
(note that the matrix $\hat g$ is not symmetric, in general). Then, the superconducting critical temperature $T_c$ is found from the equation
$\det(\hat\tau_0+\hat g\hat M)=0$, where
\begin{eqnarray}
\label{M}
    M_{\lambda\lambda'}=-\delta_{\lambda\lambda'}
    \ln\frac{2e^{C}\omega_c}{\pi T_c}+\left(\delta_{\lambda\lambda'}-\frac{\rho_{\lambda'}}{2}
    \langle\phi_{\lambda}\rangle\langle\phi_{\lambda'}\rangle\right)\mathcal{F}\left(\frac{T_c}{\Gamma}\right),
\end{eqnarray}
see Ref. \cite{MS07}.

In the absence of impurities, the second term in $M_{\lambda\lambda'}$ vanishes, and we obtain the critical temperature of a clean
superconductor:
\begin{equation}
\label{Tc0}
    T_{c0}=\frac{2e^{C}\omega_c}{\pi}e^{-1/g},
\end{equation}
where
\begin{equation}
\label{g def}
    g=\frac{g_{++}+g_{--}}{2}+\sqrt{\left(\frac{g_{++}-g_{--}}{2}\right)^2+g_{+-}g_{-+}}
\end{equation}
is the effective coupling constant. In the presence of impurities, the cases of conventional and
unconventional pairing have to be considered separately.

\underline{Unconventional pairing}. 
In this case $\langle\phi_{\lambda}\rangle=0$, and we obtain the following equation for $T_c$:
\begin{equation}
\label{Tc zero B uncon}
    \ln\frac{T_{c0}}{T_c}=\mathcal{F}\left(\frac{T_c}{\Gamma}\right).
\end{equation}
The reduction of the critical temperature is described by a universal function, which has the same form as in 
isotropic centrosymmetric superconductors with magnetic impurities,\cite{AG60} or in anisotropically paired centrosymmetric superconductors
with nonmagnetic impurities.\cite{Lar65,Book} In particular, at weak disorder, i.e. in the limit $\Gamma\ll T_{c0}$, we have
$T_c=T_{c0}-\pi\Gamma/4$. The superconductivity is completely suppressed at $\Gamma_c=(\pi/2e^C)T_{c0}$.

\underline{Conventional pairing}. Assuming a completely isotropic pairing with $\phi_{\lambda}=1$, we obtain:
\begin{eqnarray}
\label{Tc zero B con}
    \ln\frac{T_{c0}}{T_c}=\frac{1+c_1\mathcal{F}(x)}{
    c_2+c_3\mathcal{F}(x)+\sqrt{c_4+c_5\mathcal{F}(x)
    +c_6\mathcal{F}^2(x)}}-\frac{1}{g},
\end{eqnarray}
where $x=T_c/\Gamma$, and
\begin{eqnarray*}
    &&c_1=\frac{\rho_+(g_{--}-g_{+-})+\rho_-(g_{++}-g_{-+})}{2},\quad c_2= \frac{g_{++}+g_{--}}{2},\quad c_3=\frac{\det\hat g}{2},\\
    &&c_4=\left(\frac{g_{++}-g_{--}}{2}\right)^2+g_{+-}g_{-+},\quad 
	c_5=(c_2-c_1)\det\hat g,\quad c_6=c_3^2.
\end{eqnarray*}
We see that the critical temperature depends on nonmagnetic
disorder, but in contrast to the unconventional case, the effect
is not described by a universal Abrikosov-Gor'kov function.\cite{FAMS06,MS07} At
weak disorder the suppression is linear in the scattering rate, but with a non-universal slope:
\begin{equation}
\label{Tc weak disorder con}
    T_c=T_{c0}-\frac{1}{g}\left[c_1-\frac{\pi}{4}\frac{1}{g}\left(c_3+
    \frac{c_5}{2\sqrt{c_4}}\right)\right]\Gamma.
\end{equation}
In the case of strong impurity scattering, $\Gamma\gg T_{c0}$, we use $\mathcal{F}(x)=\ln(1/x)+O(1)$ 
at $x\to 0$, to find that the critical temperature approaches the limiting value given by
\begin{equation}
\label{Tc limit}
    T_c^*=T_{c0}\exp\left(\frac{1}{g}-\frac{c_1}{2c_3}\right),
\end{equation}
i.e. superconductivity is not completely destroyed by impurities.
The explanation is the same as in the conventional two-gap
superconductors, see e.g. Refs. \cite{MP66,Kusa70,Gol97}:
Interband impurity scattering tends to reduce the difference
between the gap magnitudes in the two bands, which costs energy
and thus suppresses $T_c$, but only until both gaps become equal.
One can show that both the coefficient in front of $\Gamma$ in
Eq. (\ref{Tc weak disorder con}) and the exponent in Eq. (\ref{Tc
limit}) are negative, i.e. $T_c^*<T_{c0}$.

In the BCS-like model (\ref{H-BCS}) the pairing is isotropic and described by a single coupling constant
$V_{\lambda\lambda'}=V/2$, and we have $g=N_FV$. Although the
expression (\ref{Tc0}) for the critical temperature in the clean case has the usual
BCS form, the analogy is not complete, because the order
parameter resides in two nondegenerate bands, and $N_F=(N_++N_-)/2$.
In the presence of impurities, the right-hand side of Eq.
(\ref{Tc zero B con}) vanishes, therefore there is an analog
of Anderson's theorem: The nonmagnetic disorder has no effect on the critical temperature. 

Another important particular case, possibly relevant to CePt$_3$Si, is the model in which only one, say $\lambda=+$, band is
superconducting, while the other band remains normal.\cite{SZB04,MS07} This can be described by setting
$V_{+-}=V_{--}=0$. Using Eq. (\ref{M}), we obtain the following equation for the
critical temperature:
\begin{equation}
\label{Tc-one-band}
    \ln\frac{T_{c0}}{T_c}=c_0\mathcal{F}\left(\frac{T_c}{\Gamma}\right),
\end{equation}
where $c_0=1-\rho_+\langle\phi_+\rangle^2/2$. At weak disorder we have
$T_c=T_{c0}-c_0(\pi\Gamma/4)$, while at strong disorder 
$T_c=T_{c0}(\pi T_{c0}/e^C\Gamma)^{1/c_0}$. If the pairing is anisotropic but conventional, then, unlike the unconventional case with $\langle\phi_+\rangle=0$, 
the superconductivity is never completely destroyed, even at strong disorder.

\subsection{Isotropic model}
\label{sec: isotropic model}

Finding the superconducting gap at arbitrary temperatures and impurity concentrations from the nonlinear gap equations (\ref{gap-eq gen}) is more difficult
than the calculation of $T_c$.
We focus on the case when the pairing is completely isotropic, i.e. $\phi_+(\bk)=\phi_-(\bk)=1$ and
$\tilde\Delta_\lambda(\bk)=\eta_\lambda$. The order parameter components can be chosen to be real, and the gap equations take the following form:
\begin{equation}
\label{gap-eq-isotropic}
	\sum_{\lambda'}V^{-1}_{\lambda\lambda'}\eta_{\lambda'}=\pi\rho_\lambda N_F T\sum_n{}'
    \frac{D_\lambda(\omega_n)}{\sqrt{\tilde\omega_n^2+D_\lambda^2(\omega_n)}}.
\end{equation}
We further assume that the difference between $\rho_+$ and $\rho_-$ can be neglected and the pairing
strength, see Eq. (\ref{pairing potential}), does not vary between
the bands: $V_{++}=V_{--}>0$. For the interband coupling constants, we have $V_{+-}=V_{-+}$. The gap equations have two solutions: $\eta_+=\eta_-=\eta$ and
$\eta_+=-\eta_-=\eta$. In the spin representation, the former corresponds to the singlet state, while the latter -- to the ``protected''
triplet state.\cite{FAKS04} The impurity responses of these two states turn out to be very different.

\underline{$\eta_+=\eta_-=\eta$}. In this case $D_+(\omega_n)=D_-(\omega_n)=D(\omega_n)$, and Eqs. (\ref{tilde omega
eq}) and (\ref{D eq uniform}) take the following form:
\begin{eqnarray*}
    \tilde\omega_n=\omega_n+\Gamma\frac{\tilde\omega_n}{\sqrt{\tilde\omega_n^2+D^2}},\quad 
	D=\eta+\Gamma\frac{D}{\sqrt{\tilde\omega_n^2+D^2}}.
\end{eqnarray*}
Introducing $Z(\omega_n)=1+\Gamma/\sqrt{\omega_n^2+\eta^2}$, the solution of these equations is $D(\omega_n)=Z(\omega_n)\eta$,
$\tilde\omega_n=Z(\omega_n)\omega_n$. Therefore, the gap equation (\ref{gap-eq-isotropic}) becomes
\begin{equation}
\label{gap eq singlet}
    \eta=\pi g_1T\sum_n{}'\frac{D}{\sqrt{\tilde\omega_n^2+D^2}}=
    \pi g_1T\sum_n{}'\frac{\eta}{\sqrt{\omega_n^2+\eta^2}},
\end{equation}
where $g_1=(V_{++}+V_{+-})N_F$. The scattering rate has dropped out, therefore there is an analog of Anderson's theorem:
neither the gap magnitude nor the critical temperature,
are affected by impurities. Namely, we have $T_c(\Gamma)=T_{c0}$, see Eq. (\ref{Tc0}) with $g=g_1$, while the
gap magnitude at $T=0$ is given by the clean BCS expression:
$\eta(T=0)=\eta_0=(\pi/e^C)T_{c0}$.

\underline{$\eta_+=-\eta_-=\eta$}. In this case $D_+(\omega_n)=-D_-(\omega_n)=\eta$, and we obtain from Eqs. (\ref{tilde
omega eq}), (\ref{D eq uniform}), and (\ref{gap-eq-isotropic}):
\begin{equation}
\label{eqs triplet}
    \tilde\omega_n=\omega_n+\Gamma
    \frac{\tilde\omega_n}{\sqrt{\tilde\omega_n^2+\eta^2}},\quad
	\eta=\pi g_2 T\sum_n{}'\frac{\eta}{\sqrt{\tilde\omega_n^2+\eta^2}},
\end{equation}
where $g_2=(V_{++}-V_{+-})N_F$. In the absence of impurities, the
critical temperature is given by the BCS expression (\ref{Tc0}) with $g=g_2$. If
$V_{+-}>0$ (attractive interband interaction), then $g_2<g_1$ and
the phase transition occurs into the state
$\eta_+=\eta_-$. However, if $V_{+-}<0$ (repulsive interband interaction),
then $g_2>g_1$ and the phase transition
occurs into the state $\eta_+=-\eta_-$.
In contrast to the previous case, both the critical temperature
and the gap magnitude are now suppressed by disorder. The former is determined by the equation (\ref{Tc zero B con}), which takes the same universal 
form as the Abrikosov-Gor'kov equation (\ref{Tc zero B uncon}).
Superconductivity is completely destroyed if the disorder strength exceeds the critical
value $\Gamma_c=(\pi/2e^C)T_{c0}$.

To find the gap magnitude at $T=0$ as a function of $\Gamma$ we follow the procedure
described in Ref. \cite{MK01}. Replacing the Matsubara sum by a frequency integral in the second of equations (\ref{eqs triplet}), we obtain:
\begin{equation}
\label{ln eta}
   \ln\frac{\eta_0}{\eta}=\int_0^\infty d\omega\left(\frac{1}{\sqrt{\omega^2+\eta^2}}
    -\frac{1}{\sqrt{\tilde\omega^2+\eta^2}}\right),
\end{equation}
where $\eta_0=2\Gamma_c$ is the BCS gap magnitude in the clean case, and $\tilde\omega$ satisfies the equation
$\tilde\omega=\omega+\Gamma\tilde\omega/\sqrt{\tilde\omega^2+\eta^2}$. Transforming the
second term on the right-hand side of Eq. (\ref{ln eta}) into an integral over $\tilde\omega$ we arrive at the following equation:
\begin{eqnarray}
\label{eq for eta}
    \ln\frac{\Gamma_c}{\Gamma}=\frac{\pi x}{4}-\ln(2x)+\theta(x-1)\biggl[\ln(x+\sqrt{x^2-1})\nonumber\\
    -\frac{x}{2}\arctan\sqrt{x^2-1}-\frac{\sqrt{x^2-1}}{2x}\biggr],
\end{eqnarray}
where $x=\Gamma/\eta$.
This equation does not have solutions at $\Gamma>\Gamma_c$, which is consistent with the complete
suppression of superconductivity above the critical disorder strength. In the
weak disorder limit, $\Gamma\ll\Gamma_c$, the solution is $x\simeq\Gamma/2\Gamma_c$, while  
at $\Gamma\to\Gamma_c$ we have $x\simeq\sqrt{\Gamma_c/12(\Gamma_c-\Gamma)}$.

\section{Upper critical field at arbitrary temperature}
\label{sec: Hc2}

In this section, we calculate the upper critical field $H_{c2}(T)$ of a disordered noncentrosymmetric superconductor described by the BCS-like model (\ref{H-BCS}).
We assume a uniform external field $\bH$ and neglect the paramagnetic pair breaking. The noninteracting part of the Hamiltonian is given by 
\begin{equation}
\label{hat h}
    \hat{h}=\epsilon_0(\bK)+\bgam(\bK)\bsigma+U(\br),
\end{equation}
where $\bK=-i\bm{\nabla}+(e/c)\bA(\br)$, $\bA$ is the vector potential, and $e$ is the absolute
value of the electron charge. The superconducting order parameter in the model (\ref{H-BCS}) is represented by a single complex
function $\eta(\br)$. According to Sect. \ref{sec: Tc}, the zero-field critical temperature is not affected by scalar impurities. The critical temperature at $\bH\neq 0$, or inversely the upper critical field as a function of temperature, can be found from the condition that the linearized gap equation
$[V^{-1}-T\sum_n{}^\prime\hat X(\omega_n)]\eta(\br)=0$ has a nontrivial solution. Here the operator $\hat X(\omega_n)$ is defined by the kernel
\begin{equation}
\label{X def}
    X(\br,\br';\omega_n)=\frac{1}{2}\bigl\langle\tr\hat g^\dagger\hat G(\br,\br';\omega_n)
    \hat g\hat G^T(\br,\br';-\omega_n)\bigr\rangle_{imp},
\end{equation}
where $\hat g=i\hat\sigma_2$. The angular brackets denote the impurity averaging, and
$\hat G(\br,\br';\omega_n)$ is the Matsubara Green's functions of
electrons in the normal state, which satisfies the equation
$(i\omega_n-\hat{h})\hat G(\br,\br';\omega_n)=\delta(\br-\br')$, with $\hat{h}$ given by expression (\ref{hat h}).
 
\begin{figure}[t]
    \includegraphics[width=8.1cm]{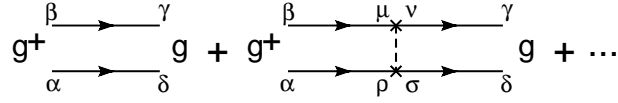}
    \caption{Impurity ladder diagrams in the Cooper channel. Lines with
    arrows correspond to the average Green's functions of electrons,
    $\hat g=i\hat\sigma_2$, and the impurity (dashed) line
    is defined in the text, see Eq. (\ref{impurity line}).}
    \label{fig: ladder diagrams}
\end{figure}

The impurity average of the product of two Green's functions in Eq. (\ref{X def}) can be represented graphically by the ladder
diagrams, see Fig. \ref{fig: ladder diagrams} (as before, we assume the disorder to be sufficiently weak for the diagrams with crossed
impurity lines to be negligible). In order to sum the diagrams, we introduce an impurity-renormalized gap function $\hat
D(\br,\omega_n)$, which a matrix in the spin space satisfying the following integral equation:
\begin{eqnarray}
\label{D eq}
    &&\hat D(\br,\omega_n)=\eta(\br)\hat g\nonumber\\
    &&\quad+\frac{1}{2}n_{imp}U_0^2\hat g
    \int d^3\br'\tr\hat g^\dagger
    \langle\hat G(\br,\br';\omega_n)\rangle_{imp}\hat D(\br',\omega_n)\langle\hat G^T(\br,\br';-\omega_n)\rangle_{imp}\nonumber\\
    &&\quad+\frac{1}{2}n_{imp}U_0^2\hat{\bm{g}}\int d^3\br'\tr\hat{\bm{g}}^\dagger
    \langle\hat G(\br,\br';\omega_n)\rangle_{imp}\hat D(\br',\omega_n)\langle\hat G^T(\br,\br';-\omega_n)\rangle_{imp}.\quad
\end{eqnarray}
This can be easily derived from the ladder diagrams in Fig. \ref{fig:
ladder diagrams}, by representing each ``rung'' of the ladder as a
sum of spin-singlet and spin-triplet terms:
\begin{equation}
\label{impurity line}
    n_{imp}U_0^2\delta_{\mu\nu}\delta_{\rho\sigma}=
    \frac{1}{2}n_{imp}U_0^2g_{\mu\rho}g^\dagger_{\sigma\nu}+
    \frac{1}{2}n_{imp}U_0^2\bm{g}_{\mu\rho}\bm{g}^\dagger_{\sigma\nu},
\end{equation}
where $\hat{\bm{g}}=i\hat{\bm{\sigma}}\hat\sigma_2$.

Seeking solution of Eq. (\ref{D eq}) in the form $\hat D(\br,\omega_n)=d_0(\br,\omega_n)\hat g+\bm{d}(\br,\omega_n)\hat{\bm{g}}$,
we obtain a system of four integral equations for $d_a$, where $a=0,1,2,3$:
\begin{equation}
\label{dd eqs}
    \sum_{b=0}^3\bigl[\delta_{ab}-\Gamma\hat{\mathcal{Y}}_{ab}(\omega_n)\bigr]
    d_b(\br,\omega_n)=\eta(\br)\delta_{a0}.
\end{equation}
Here the operators $\hat{\mathcal{Y}}_{ab}(\omega_n)$ are defined
by the kernels
\begin{equation}
\label{Yab def}
    \mathcal{Y}_{ab}(\br,\br';\omega_n)=\frac{1}{2\pi N_F}\tr\hat{\mathrm{g}}_a^\dagger
    \langle\hat G(\br,\br';\omega_n)\rangle_{imp}\hat{\mathrm{g}}_b\langle\hat G^T(\br,\br';-\omega_n)\rangle_{imp},
\end{equation}
with $\hat{\mathrm{g}}_0=\hat g$, and $\hat{\mathrm{g}}_a=\hat
g_a$ for $a=1,2,3$. We see that, in addition to the spin-singlet
component $d_0(\br,\omega_n)$, impurity scattering also induces a nonzero spin-triplet component $\bm{d}(\br,\omega_n)$.
The linearized gap equation contains only the former. Indeed, using Eqs. (\ref{dd eqs}) we obtain:
\begin{equation}
\label{gap eq d0}
    \frac{1}{N_FV}\eta(\br)-\pi T\sum_n{}'\frac{d_0(\br,\omega_n)-\eta(\br)}{\Gamma}=0.
\end{equation}
It is easy to see that the triplet component does not appear
in the centrosymmetric case. Indeed, in the absence of the Zeeman
interaction the spin structure of the Green's function is trivial:
$G_{\alpha\beta}(\br,\br';\omega_n)=\delta_{\alpha\beta}G(\br,\br';\omega_n)$.
Then it follows from Eq. (\ref{Yab def}) that
$\hat{\mathcal{Y}}_{ab}(\omega_n)=\delta_{ab}\hat{\mathcal{Y}}(\omega_n)$,
therefore $d_0=(1-\Gamma\hat{\mathcal{Y}})^{-1}\eta$ and
$\bm{d}=0$.

The next step is to find the spectrum of the operators
$\hat{\mathcal{Y}}_{ab}(\omega_n)$. At zero field, the average Green's function has the following form:
\begin{equation}
\label{G zero H}
    \hat{G}_0(\bk,\omega_n)=
    \sum_{\lambda=\pm}\hat\Pi_\lambda(\bk)G_\lambda(\bk,\omega_n),
\end{equation}
where $\hat\Pi_\lambda=(1+\lambda\hat{\bgam}\bsigma)/2$
are the helicity band projection operators, and $G_\lambda(\bk,\omega_n)$
are the impurity-averaged Green's functions in the band representation, see Eq. (\ref{GF_n}). 
At $\bH\neq 0$, we have $\langle\hat G(\br,\br';\omega_n)\rangle_{imp}=\hat
G_0(\br-\br';\omega_n)\exp[(ie/c)\int_{\br}^{\br'}\bm{A}d\br]$, where the
integration is performed along a straight line connecting $\br$
and $\br'$.\cite{Gorkov59} This approximation is legitimate if the temperature is not very low, so
that the Landau level quantization can be neglected. It follows from Eq. (\ref{Yab def}) that
$\hat{\mathcal{Y}}_{ab}(\omega_n)=Y_{ab}(\bq,\omega_n)\bigr|_{\bq\to\bD}$, where
$\bD=-i\bnab+(2e/c)\bm{A}$ and
\begin{equation}
\label{Yab}
    Y_{ab}(\bq,\omega_n)=\frac{1}{2\pi N_F}\int\frac{d^3\bk}{(2\pi)^3}\tr
    \hat{\mathrm{g}}_a^\dagger\hat G_0(\bk+\bq,\omega_n)
    \hat{\mathrm{g}}_b\hat G_0^T(-\bk,-\omega_n).
\end{equation}
Substituting here expressions (\ref{G zero H}) and calculating the spin traces, we obtain for the 
singlet-singlet and singlet-triplet contributions:
\begin{eqnarray}
\label{Y00}
    &&Y_{00}=\frac{1}{2}\sum_\lambda\rho_\lambda
    \left\langle\frac{1}{|\omega_n|+\Gamma+i\bm{v}_\lambda(\bk)\bq\sign\omega_n/2}
    \right\rangle,\\
\label{Y0i}
    &&Y_{0i}=Y_{i0}
    =\frac{1}{2}\sum_\lambda\lambda\rho_\lambda\left\langle\frac{\hat\gamma_i(\bk)}{|\omega_n|
    +\Gamma+i\bm{v}_\lambda(\bk)\bq\sign\omega_n/2}\right\rangle,
\end{eqnarray}
where $\bm{v}_\lambda=\partial\xi_\lambda/\partial\bk$ is the quasiparticle velocity in the $\lambda$th band.
We see that the singlet-triplet mixing occurs due to the SO coupling and vanishes
when $\rho_+=\rho_-=1$ and $\bm{v}_+=\bm{v}_-=\bm{v}_F$. The triplet-triplet contributions can be represented as follows:
$Y_{ij}=Y^{(1)}_{ij}+Y^{(2)}_{ij}$, where
\begin{equation}
\label{Y1ij}
    Y^{(1)}_{ij}=\frac{1}{2}\sum_\lambda\rho_\lambda\left\langle
    \frac{\hat\gamma_i(\bk)\hat\gamma_j(\bk)}{|\omega_n|+\Gamma+i\bm{v}_\lambda(\bk)\bq\sign\omega_n/2}
    \right\rangle,
\end{equation}
and
\begin{equation}
\label{Y2ij}
    Y^{(2)}_{ij}=\frac{1}{2\pi N_F}\sum_\lambda\int\frac{d^3\bk}{(2\pi)^3}
    (\delta_{ij}-\hat\gamma_i\hat\gamma_j-i\lambda e_{ijl}\hat\gamma_l)
    G_\lambda(\bk+\bq,\omega_n)G_{-\lambda}(-\bk,-\omega_n).
\end{equation}
The singlet impurity scattering channel, which is described by the
first term in Eq. (\ref{impurity line}), causes only the
scattering of intraband pairs between the bands. In contrast, the
triplet impurity scattering can also create interband pairs, which
are described by $Y^{(2)}_{ij}$. 

It is easy to show that if the SO band splitting exceeds both $\omega_c$ and
$\Gamma$, then the interband term in $Y_{ij}$ is
smaller than the intraband one. Let us consider, for example, isotropic
bands with $\xi_\pm(\bk)=\epsilon_0(\bk)\pm\gamma$. Neglecting
for simplicity the differences between the densities of states and
the Fermi velocities in the two bands and setting $\bq=0$, we obtain from
Eqs. (\ref{Y1ij}) and (\ref{Y2ij}):
\begin{eqnarray*}
    &&Y^{(1)}_{ij}(\bm{0},\omega_n)
    =\frac{\delta_{ij}}{3(|\omega_n|+\Gamma)}\equiv Y_{intra}(\omega_n)\delta_{ij},\\
    &&Y^{(2)}_{ij}(\bm{0},\omega_n)=
    \frac{2\delta_{ij}}{3(|\omega_n|+\Gamma)(1+r^2)}\equiv Y_{inter}(\omega_n)\delta_{ij},
\end{eqnarray*}
where $r(\omega_n)=\gamma/(|\omega_n|+\Gamma)$. Due to the BCS
cutoff, the maximum value of $\omega_n$ in the Cooper ladder is equal to $\omega_c$,
therefore $r_{min}\sim E_{SO}/\max(\omega_c,\Gamma)$. We assume that this ratio is large, which is a good assumption for real materials, 
therefore
\begin{equation}
\label{inter-to-intra}
    \max_n\frac{Y_{inter}(\omega_n)}{Y_{intra}(\omega_n)}=\frac{2}{1+r_{min}^2}\sim
    \left[\frac{\max(\omega_c,\Gamma)}{E_{SO}}\right]^2\ll 1, 
\end{equation}
at all Matsubara frequencies satisfying $|\omega_n|\leq\omega_c$.

Thus the interband contributions to the Cooper ladder can be neglected, and we obtain:
\begin{equation}
\label{Yab intraband}
    Y_{ab}(\bq,\omega_n)=\frac{1}{2}\sum_\lambda\rho_\lambda\left\langle
    \frac{\Lambda_{\lambda,a}(\bk)\Lambda_{\lambda,b}(\bk)}{|\omega_n|+\Gamma
    +i\bm{v}_\lambda(\bk)\bq\sign\omega_n/2}\right\rangle,
\end{equation}
where $\Lambda_{\lambda,0}(\bk)=1$ and $\Lambda_{\lambda,a}(\bk)=\lambda\hat\gamma_a(\bk)$ for $a=1,2,3$.
Making the substitution $\bq\to\bD$, we represent $\hat{\mathcal{Y}}_{ab}$ as a differential operator of
infinite order:
\begin{equation}
\label{hat Yab final}
    \hat{\mathcal{Y}}_{ab}(\omega_n)=\frac{1}{2}\int_0^\infty du\;e^{-u(|\omega_n|+\Gamma)}
    \sum_\lambda\rho_\lambda\left\langle\Lambda_{\lambda,a}(\bk)\Lambda_{\lambda,b}(\bk)
    e^{-iu\bm{v}_\lambda(\bk)\bD\sign\omega_n/2}\right\rangle.
\end{equation}

In order to solve Eqs. (\ref{dd eqs}), with the operators
$\hat{\mathcal{Y}}_{ab}(\omega_n)$ given by expressions (\ref{hat
Yab final}), we follow the procedure described in Ref.
\cite{HW66}. Choosing the $z$-axis along the external
field: $\bH=H\hat z$, we introduce the operators
$a_\pm=\ell_H(D_x\pm iD_y)/2$, and $a_3=\ell_HD_z$,
where $\ell_H=\sqrt{c/eH}$ is the magnetic length. It is easy to
check that $a_+=a_-^\dagger$ and $[a_-,a_+]=1$, therefore $a_\pm$
have the meaning of the raising and lowering operators, while
$a_3=a_3^\dagger$ commutes with both of them: $[a_3,a_\pm]=0$. It
is convenient to expand both the order parameter $\eta$ and the
impurity-renormalized gap functions $d_a$ in the basis of Landau
levels $|N,p\rangle$, which satisfy
$a_+|N,p\rangle=\sqrt{N+1}|N+1,p\rangle$, $a_-|N,p\rangle=\sqrt{N}|N-1,p\rangle$, and
$a_3|N,p\rangle=p|N,p\rangle$, where $N=0,1,...$, and $p$ is a real number characterizing the variation of the order parameter 
along the field. We have
\begin{equation}
\label{eta expand}
    \eta(\br)=\sum_{N,p}\eta_{N,p}\langle\br|N,p\rangle,\qquad
    d_a(\br,\omega_n)=\sum_{N,p}d^a_{N,p}(\omega_n)\langle\br|N,p\rangle.
\end{equation}
According to Eqs. (\ref{dd eqs}), the expansion coefficients can be found from the following algebraic equations:
\begin{equation}
\label{dd eqs linear}
    \sum_{N',p',b}\Bigl[\delta_{ab}\delta_{NN'}\delta_{pp'}
    -\Gamma\langle N,p|\hat{\mathcal{Y}}_{ab}(\omega_n)|N',p'\rangle\Bigr]
    d^b_{N',p'}(\omega_n)=\delta_{a0}\eta_{N,p}.
\end{equation}
Substituting the solutions of these equations into
\begin{equation}
\label{gap eq LL}
    \frac{1}{N_FV}\eta_{N,p}-\pi T\sum_n{}'\frac{d^0_{N,p}(\omega_n)-\eta_{N,p}}{\Gamma}=0,
\end{equation}
see Eq. (\ref{gap eq d0}), and setting the determinant of the
resulting linear equations for $\eta_{N,p}$ to zero, one arrives
at an equation for the upper critical field.

\subsection{$H_{c2}(T)$ in a cubic crystal}
\label{sec: Hc2 cubic}

In the general case, i.e. for arbitrary crystal symmetry and electronic band structure, the procedure outlined above
does not yield an equation for $H_{c2}(T)$ in a closed form, since all the Landau levels are coupled.
In order to make progress, we focus on the cubic case, $\mathbf{G}=\mathbf{O}$, with a parabolic band and the SO coupling given by $\bgam(\bk)=\gamma_0\bk$. 
As for the parameter $\delta$, which characterizes the difference between the band densities of states, see Eq. (\ref{rhos}), we assume that
\begin{equation}
\label{delta range}
    \delta_c\ll|\delta|\leq 1,
\end{equation}
where $\delta_c=\max(\omega_c,\Gamma)/\epsilon_F\ll 1$. While the
first inequality follows from the condition
(\ref{inter-to-intra}), which ensures the smallness of
the interband contribution to the Cooper impurity ladder, the second one is always satisfied,
with $|\delta|\to 1$ corresponding to the rather unrealistic limit of extremely strong SO coupling.

In order to solve the gap equations, we make a change of variables
in the triplet component: $d_\pm=(d_1\pm id_2)/\sqrt{2}$.
Then, Eqs. (\ref{dd eqs}) take the following form:
\begin{equation}
\label{dd eqs cubic}
    \left( \begin{array}{cccc}
    1-\Gamma\hat{\mathcal{Y}}_{00} & -\Gamma\hat{\mathcal{Y}}_{03} & -\Gamma\hat{\mathcal{Y}}_{0-} & -\Gamma\hat{\mathcal{Y}}_{0+} \\
    -\Gamma\hat{\mathcal{Y}}_{03} & 1-\Gamma\hat{\mathcal{Y}}_{33} & -\Gamma\hat{\mathcal{Y}}_{3-} & -\Gamma\hat{\mathcal{Y}}_{3+} \\
    -\Gamma\hat{\mathcal{Y}}_{0+} & -\Gamma\hat{\mathcal{Y}}_{3+} & 1-\Gamma\hat{\mathcal{Z}} & -\Gamma\hat{\mathcal{Z}}_{+} \\
    -\Gamma\hat{\mathcal{Y}}_{0-} & -\Gamma\hat{\mathcal{Y}}_{3-} & -\Gamma\hat{\mathcal{Z}}_{-} & 1-\Gamma\hat{\mathcal{Z}}
    \end{array} \right)
    \left( \begin{array}{c}
    d_0 \\ d_3 \\ d_+ \\ d_-
    \end{array} \right)
    =\left( \begin{array}{c}
    \eta \\ 0 \\ 0 \\ 0
    \end{array} \right),
\end{equation}
where
\begin{equation}
\label{YZ-ops}
	\begin{array}{l}
    \displaystyle \hat{\mathcal{Y}}_{0\pm}=\frac{\hat{\mathcal{Y}}_{01}\pm i\hat{\mathcal{Y}}_{02}}{\sqrt{2}}, \qquad
    \displaystyle \hat{\mathcal{Y}}_{3\pm}=\frac{\hat{\mathcal{Y}}_{13}\pm i\hat{\mathcal{Y}}_{23}}{\sqrt{2}}, \\
    \displaystyle \hat{\mathcal{Z}}=\frac{\hat{\mathcal{Y}}_{11}+\hat{\mathcal{Y}}_{22}}{2}, \qquad
    \displaystyle \hat{\mathcal{Z}}_{\pm}=\frac{\hat{\mathcal{Y}}_{11}\pm 2i\hat{\mathcal{Y}}_{12}-\hat{\mathcal{Y}}_{22}}{2},
	\end{array}
\end{equation}
with $\hat{\mathcal{Y}}_{ab}=\hat{\mathcal{Y}}_{ba}$ given by Eqs. (\ref{hat Yab final}).

According to Eq. (\ref{dd eqs linear}), one has to know the matrix elements of the operators
$\hat{\mathcal{Y}}_{ab}(\omega_n)$ in the basis of the Landau
levels $|N,p\rangle$. After some straightforward algebra, see Ref. \cite{Sam-Hc2}
for details, we find that $\hat{\mathcal{Y}}_{00}$, $\hat{\mathcal{Y}}_{03}$, $\hat{\mathcal{Y}}_{33}$,
and $\hat{\mathcal{Z}}$ are diagonal in the Landau levels, while for the rest of the operators (\ref{YZ-ops}) the only nonzero matrix elements are
as follows: $\langle N,p|\hat{\mathcal{Y}}_{0-}|N+1,p\rangle=\langle N+1,p|\hat{\mathcal{Y}}_{0+}|N,p\rangle$, 
$\langle N,p|\hat{\mathcal{Y}}_{3-}|N+1,p\rangle=\langle N+1,p|\hat{\mathcal{Y}}_{3+}|N,p\rangle$, and 
$\langle N,p|\hat{\mathcal{Z}}_-|N+2,p\rangle=\langle N+2,p|\hat{\mathcal{Z}}_+|N,p\rangle$.
Therefore, the Landau levels are decoupled, and for $\eta(\br)=\eta\langle\br|N,p\rangle$ ($\eta$
is a constant) the solution of Eqs. (\ref{dd eqs cubic}) has the
following form:
\begin{equation}
\label{d solution cubic}
    \left(\begin{array}{c}
    d_0(\br,\omega_n)\\
    d_3(\br,\omega_n)\\
    d_+(\br,\omega_n)\\
    d_-(\br,\omega_n)
    \end{array} \right)
    =\left(\begin{array}{c}
    d^0_{N,p}(\omega_n)\langle\br|N,p\rangle\\
    d^3_{N,p}(\omega_n)\langle\br|N,p\rangle\\
    d^+_{N,p}(\omega_n)\langle\br|N+1,p\rangle\\
    d^-_{N,p}(\omega_n)\langle\br|N-1,p\rangle
    \end{array} \right).
\end{equation}

At arbitrary magnitude of the SO band splitting, the singlet-triplet mixing makes the equation for $H_{c2}(T)$ in
noncentrosymmetric superconductors considerably more cumbersome
than in the Helfand-Werthamer problem,\cite{HW66} even in our ``minimal''
isotropic model. It is even possible that, at some values of the
parameters, the maximum critical field is achieved for $N>0$ and
$p\neq 0$, the latter corresponding to a disorder-induced
modulation of the order parameter along the applied field. Leaving these exotic possibilities aside, here
we just assume that $N=p=0$. Then the only nonzero components of the impurity-renormalized 
gap function, see Eq. (\ref{eta expand}), are $d^0_{0,0}$ and $d^+_{0,0}$.

It is convenient to introduce the reduced temperature, magnetic field, and disorder strength:
$$
    t=\frac{T}{T_{c0}},\qquad h=\frac{2H}{H_0},\qquad
    \zeta=\frac{\Gamma}{\pi T_{c0}},
$$
where $H_0=\Phi_0/\pi\xi_0^2$, $\Phi_0=\pi c/e$ is the magnetic flux quantum, and $\xi_0=v_F/2\pi T_{c0}$ is the superconducting
coherence length ($v_F$ is the Fermi velocity). Using the expression for the critical temperature to eliminate both 
the frequency cutoff and the coupling constant from Eq. (\ref{gap eq LL}), we
arrive at the following equation for the upper critical field $h_{c2}(t)$:
\begin{equation}
\label{hc2 eq reduced}
    \ln\frac{1}{t}=2\sum_{n\geq 0}\left[\frac{1}{2n+1}-t
    \frac{w_n(1-\zeta p_n)-\zeta\delta^2q_n^2}{(1-\zeta w_n)(1-\zeta p_n)+\zeta^2\delta^2q_n^2}\right],
\end{equation}
where
\begin{eqnarray}
\label{w p q}
        &&w_n=\int_0^\infty d\rho\,e^{-\theta_n\rho}\int_0^1 ds\,
    e^{-h\rho^2(1-s^2)/4},\nonumber\\
    &&p_n=\int_0^\infty d\rho\,e^{-\theta_n\rho}\int_0^1 ds\,
    \frac{1-s^2}{2}\left[1-\frac{h}{2}\rho^2(1-s^2)\right]e^{-h\rho^2(1-s^2)/4},\\
    &&q_n=\int_0^\infty d\rho\,e^{-\theta_n\rho}\int_0^1 ds\,
    \sqrt{\frac{h}{4}}\rho(1-s^2)e^{-h\rho^2(1-s^2)/4},\nonumber
\end{eqnarray}
where $\theta_n=(2n+1)t+\zeta$. 

In the clean limit, i.e. at $\zeta\to 0$, or if the SO band splitting is negligibly small, i.e. at $\delta\to 0$,
the Helfand-Werthamer equation for $H_{c2}$ in a centrosymmetric superconductor\cite{HW66} is recovered. Therefore, the absence of inversion
symmetry affects the upper critical field only if disorder is
present. One can expect that the effect will be most pronounced in
the ``dirty'' limit, in which Eq. (\ref{hc2 eq reduced}) takes the form of a universal equation 
describing the magnetic pair breaking in superconductors:\cite{Tink-book}
\begin{equation}
\label{universal eq}
    \ln\frac{1}{t}=\Psi\left(\frac{1}{2}+\frac{\sigma}{t}\right)
    -\Psi\left(\frac{1}{2}\right).
\end{equation}
Here the parameter
\begin{equation}
\label{pair breaking}
    \sigma=\frac{2+\delta^2}{12\zeta}h
\end{equation}
characterizes the pair-breaker strength. Note that the
corresponding expression in the centrosymmetric case is different:
$\sigma_{CS}=h/6\zeta$. Analytical expressions for the upper critical field can be obtained in the
weak-field limit near the critical temperature:
\begin{equation}
\label{hc2 dirty low h}
    h_{c2}|_{t\to 1}=\frac{24\zeta}{(2+\delta^2)\pi^2}(1-t),
\end{equation}
and also at low temperatures:
\begin{equation}
\label{hc2 dirty low t}
    h_{c2}|_{t=0}=\frac{3e^{\mathbf{-C}}}{2+\delta^2}\zeta.
\end{equation}
We see that nonmagnetic disorder suppresses the orbital pair breaking and thus enhances the upper critical field.

In the general case, $H_{c2}(T)$ can be calculated analytically only in the vicinity of the critical temperature using the Ginzburg-Landau free energy expansion. 
The results for the impurity response turn out to be nonuniversal, i.e. dependent on the pairing symmetry, the values of the intra- and interband coupling constants, 
and the densities of states in the helicity bands.\cite{MS07}

\section{Spin susceptibility}
\label{sec: susceptibility}

In this section we calculate the magnetic response of a noncentrosymmetric superconductor, neglecting the orbital magnetic interaction and 
taking into account only the Zeeman coupling of the electron spins with a uniform external field $\bH$:
\begin{eqnarray}
\label{H Zeeman}
    H_{Zeeman}=-\mu_B\bH\sum\limits_{\bk,\alpha\beta}
    \bm{\sigma}_{\alpha\beta}a^\dagger_{\bk\alpha}a_{\bk\beta}
    =-\bH\sum_{\bk,\lambda\lambda'}\bm{m}_{\lambda\lambda'}(\bk)
    c^\dagger_{\bk\lambda}c_{\bk\lambda'},
\end{eqnarray}
where $\mu_B$ is the Bohr magneton. The components of the spin magnetic moment operator in the band representation
have the following form:
\begin{eqnarray}
\label{m_i}
    &&\hat m_x=\mu_B
    \left(\begin{array}{cc}
      \hat\gamma_x & -(\gamma_x\hat\gamma_z+i\gamma_y)/\gamma_\perp \\
      -(\gamma_x\hat\gamma_z-i\gamma_y)/\gamma_\perp & -\hat\gamma_x \\
    \end{array}\right),\nonumber\\
    &&\hat m_y=\mu_B
    \left(\begin{array}{cc}
      \hat\gamma_y & -(\gamma_y\hat\gamma_z-i\gamma_x)/\gamma_\perp \\
      -(\gamma_y\hat\gamma_z+i\gamma_x)/\gamma_\perp & -\hat\gamma_y \\
    \end{array}\right),\quad\\
    &&\hat m_z=\mu_B
    \left(\begin{array}{cc}
      \hat\gamma_z & \gamma_\perp/\gamma \\
      \gamma_\perp/\gamma & -\hat\gamma_z \\
    \end{array}\right)\nonumber,
\end{eqnarray}
where $\gamma_\perp=\sqrt{\gamma_x^2+\gamma_y^2}$. 

The magnetization of the system is expressed in terms of the Green's functions as follows:
${\cal M}_i=(1/{\cal V})T\sum_n\sum_{\bk,\lambda\lambda'}\tr\,m_{i,\lambda\lambda'}(\bk){\cal G}^{11}_{\lambda'\lambda}(\bk,\omega_n)$, 
where ${\cal G}(\bk,\omega_n)$ is the impurity-averaged $4\times 4$ matrix Green's function in the presence of magnetic field (recall that the upper indices label the Nambu matrix components, see Sect. \ref{sec: SC}). In a weak field, we have ${\cal M}_i=\sum_j\chi_{ij}H_j$, where $\chi_{ij}$ is the spin susceptibility tensor. Treating the Zeeman interaction, Eq. (\ref{H Zeeman}), as a small perturbation and expanding ${\cal G}$ in powers of $\bH$, we obtain:
\begin{equation}
\label{chi-ij-gen}
	\chi_{ij}=-T\sum_n\frac{1}{\cal V}\sum_{\bk,\bk'}
	\bigl\langle\Tr M_i(\bk){\cal G}(\bk,\bk';\omega_n)M_j(\bk'){\cal G}(\bk',\bk;\omega_n)\bigr\rangle_{imp},
\end{equation}
where $M_i(\bk)=\mathrm{diag}\,[\hat{m}_i(\bk),-\hat{m}_i^T(-\bk)]$.
The Green's functions here are unaveraged $4\times 4$ matrix Green's functions at zero field, and the trace is taken in both the electron-hole and helicity 
indices.  

In the clean case, one can evaluate expression (\ref{chi-ij-gen}) by summing over the Matsubara frequencies first, followed by the momentum integration. The susceptibility tensor can be represented as $\chi_{ij}=\chi^{+}_{ij}+\chi^{-}_{ij}+\tilde\chi_{ij}$ (Ref. \cite{Sam07}), where
\begin{eqnarray}
\label{chi intra gen SC}
    \chi^{\lambda}_{ij}=-\mu_B^2T\sum_n\int\frac{d^3\bk}{(2\pi)^3}
    \hat\gamma_i\hat\gamma_j\bigl(G_\lambda^2+|\tilde F_\lambda|^2\bigr)
    =\mu_B^2N_\lambda\left\langle\hat\gamma_i\hat\gamma_jY_\lambda\right\rangle
\end{eqnarray}
are the intraband contributions, determined by the thermally-excited quasiparticles near the Fermi surfaces. Here
$Y_\lambda(\bk,T)=2\int_0^\infty d\xi(-\partial f/\partial E_\lambda)$ is the angle-resolved Yosida function, 
$f(\epsilon)=(e^{\epsilon/T}+1)^{-1}$ is the Fermi-Dirac distribution function, $E_\lambda(\bk)=\sqrt{\xi^2+|\tilde\Delta_\lambda(\bk)|^2}$ is 
the energy of quasiparticle excitations in the $\lambda$th band, and, as in the previous sections, the subscripts
$\lambda$ in the Fermi-surface averages are omitted for brevity. The interband contribution is given by
\begin{eqnarray}
\label{chi inter}
    \tilde\chi_{ij}=-2\mu_B^2T\sum_n\int\frac{d^3\bk}{(2\pi)^3}
    (\delta_{ij}-\hat\gamma_i\hat\gamma_j)\bigl(G_+G_-+\re\tilde F_+^*\tilde F_-\bigr)\nonumber\\
    \simeq-\mu_B^2\int\frac{d^3\bk}{(2\pi)^3}
    \frac{\delta_{ij}-\hat\gamma_i\hat\gamma_j}{|\bgam|}[f(\xi_+)-f(\xi_-)].
\end{eqnarray}
Since $\tilde\chi_{ij}$ is determined by all quasiparticles in the
momentum-space shell ``sandwiched'' between the Fermi surfaces, it is almost unchanged when the
system undergoes the superconducting transition, in which only the electrons near the Fermi surface are affected.

\begin{figure}
    \includegraphics[width=8.5cm]{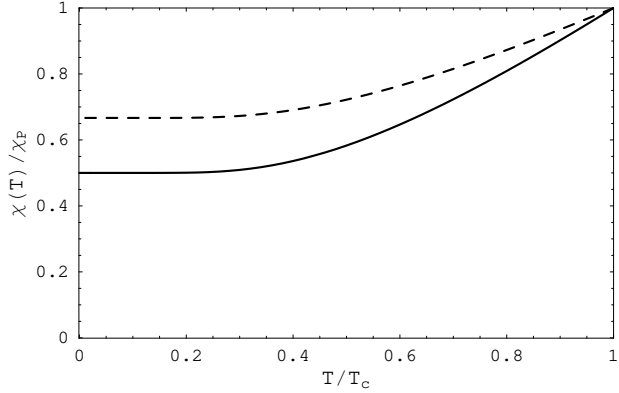}
    \caption{The clean-case temperature dependence of the transverse components of
    the susceptibility for the 2D model (the solid line), and of all
    three components for the 3D model (the dashed line); $\chi_P=2\mu_B^2N_F$ is the Pauli susceptibility.}
    \label{fig: chi clean}
\end{figure}

Collecting together the inter- and intraband contributions, we arrive at the following expression for the spin susceptibility of a clean superconductor:
\begin{equation}
\label{chi ij SC final}
    \chi_{ij}=\tilde\chi_{ij}+
    \mu_B^2N_F\sum_\lambda\rho_\lambda\left\langle\hat\gamma_i\hat\gamma_j
    Y_\lambda\right\rangle.
\end{equation}
At zero temperature there are no excitations ($Y_\lambda=0$) and the intraband terms are absent, but the susceptibility
still has a nonzero value given by $\tilde\chi_{ij}$. The temperature dependence of the susceptibility in the
superconducting state at $0<T\leq T_c$ is determined by the intraband terms, with the low-temperature
behavior depending crucially on the magnitude of the SO coupling at the gap nodes. While in the fully gapped case the
intraband susceptibility is exponentially small in all directions, in the presence of the lines of nodes it is proportional to either
$T$ or $T^3$, depending on whether or not the zeros of $\tilde\Delta_\lambda(\bk)$ coincide with those of $\bgam(\bk)$,
see Ref. \cite{Sam05}.
In Fig. \ref{fig: chi clean} the temperature dependence of $\chi_{ij}$ is plotted 
for a Rashba superconductor with $\bgam(\bk)=\gamma_\perp(\bk\times\hat z)$ and a cylindrical Fermi surface (referred 
to as the 2D model), and also for a cubic superconductor with $\bgam(\bk)=\gamma_0\bk$ and a spherical Fermi surface (the 3D model). In both 
cases, the gaps in the two helicity bands are assumed to be isotropic and have the same magnitude.

Now let us include the scalar disorder described by Eq. (\ref{H_imp}), or, equivalently, Eq. (\ref{H_imp_band}). 
After the impurity averaging, Eq. (\ref{chi-ij-gen}) is represented by a sum of the ladder diagrams containing the 
average Green's functions (\ref{GFs average}).
In contrast to the clean case, it is not possible to calculate the
Matsubara sums before the momentum integrals. To make progress, one should add to and subtract from Eq. (\ref{chi-ij-gen}) 
the normal-state susceptibility.\cite{AGD}
It is easy to show that the latter is not affected by impurities and is therefore given by $\chi_{N,ij}=\tilde\chi_{ij}+
\mu_B^2N_F\sum_\lambda\rho_\lambda\left\langle\hat\gamma_i\hat\gamma_j\right\rangle$, see Eq. (\ref{chi ij SC final}). 
Then, 
\begin{equation}
\label{delta chi}
	\chi_{ij}-\chi_{N,ij}=-T\sum_n{}'\frac{1}{\cal V}\sum_{\bk\bk'}
	\bigl(\langle\Tr M_i{\cal G}M_j{\cal G}\rangle_{imp}-\langle\Tr M_i{\cal G}_NM_j{\cal G}_N\rangle_{imp}\bigr),
\end{equation}
where ${\cal G}_N$ is the unaveraged $4\times 4$ matrix Green's function in the normal state with impurities, and the Matsubara summation is limited to the
frequencies $|\omega_n|\leq\omega_c$, at which the gap function is nonzero. 

Due to convergence of the expression on the right-hand side of Eq. (\ref{delta chi}), one can now do the momentum integrals first. 
The second term vanishes, while the calculation of the ladder diagrams in the first term is facilitated by the observation that 
the integrals of the products of the Green's functions from different bands are small compared with their counterparts containing
the Green's functions from the same band. 
The argument is similar to the one leading to Eq. (\ref{inter-to-intra}). For example, assuming that both the SO band splitting and the pairing are isotropic, 
i.e. $\xi_\pm(\bk)=\epsilon_0(\bk)\pm\gamma$ and $D_\pm=D$, we have
$$
    \max_n\frac{\int d\epsilon_0\, G_\lambda G_{-\lambda}}{\int d\epsilon_0\, G_\lambda G_{\lambda}}=\max_n\frac{1}{1+\gamma^2/\Omega_n^2}\sim
    \left[\frac{\max(\omega_c,\Gamma)}{E_{SO}}\right]^2\ll 1, 
$$
where $\Omega_n=\sqrt{\tilde\omega_n^2+|D|^2}$. In the same way, one can also obtain estimates for the momentum integrals containing anomalous Green's functions:
\begin{eqnarray*}
    &&\max_n\frac{\int d\epsilon_0\, G_\lambda \tilde F_{-\lambda}}{\int d\epsilon_0\, 
    G_\lambda\tilde F_{\lambda}}=\max_n\frac{1-i\lambda\gamma/\tilde\omega_n}{1+\gamma^2/\Omega_n^2}
    \sim\frac{\max(\omega_c,\Gamma)}{E_{SO}}\ll 1,\\
    &&\max_n\frac{\int d\epsilon_0\, \tilde F_\lambda \tilde F_{-\lambda}}{\int d\epsilon_0\, \tilde F_\lambda
    \tilde F_{\lambda}}=\max_n\frac{1}{1+\gamma^2/\Omega_n^2}\sim\left[\frac{\max(\omega_c,\Gamma)}{E_{SO}}\right]^2\ll 1. 
\end{eqnarray*}
Thus we see that it is legitimate to keep only the same-band contributions to the impurity ladder on the right-hand side of Eq. (\ref{delta chi}). 

Following Ref. \cite{Sam07}, we obtain the following expression for the spin susceptibility:
\begin{equation}
\label{chi ij impurities gen}
    \chi_{ij}=\chi_{N,ij}+\frac{2\pi\mu_BN_F}{\Gamma}T\sum_n
    \frac{\partial X_i(\omega_n)}{\partial H_j},
\end{equation}
where $X_i$ are found from the equations
\begin{equation}
\label{XY eqs gen}
    \begin{array}{l}
    \displaystyle X_i-\sum_{j}(A_{1,ij}X_j+A_{2,ij}Y_j+A^*_{2,ij}Y^*_j)
    =X_{0,i}, \\
    \displaystyle Y_i-\sum_{j}(2A^*_{2,ij}X_j+A_{3,ij}Y_j+A_{4,ij}Y^*_j)
    =Y_{0,i}. \\
    \end{array}
\end{equation}
The notations here are as follows:
\begin{eqnarray*}
    &&A_{1,ij}=\frac{\Gamma}{2}\sum_\lambda\rho_\lambda\left\langle
    \frac{\hat\gamma_i\hat\gamma_j|D_\lambda|^2}{\Omega_n^3}\right\rangle,\quad
    A_{2,ij}=\frac{\Gamma}{4}\sum_\lambda\rho_\lambda\left\langle
    \frac{i\hat\gamma_i\hat\gamma_j\tilde\omega_nD_\lambda^*}{\Omega_n^3}\right\rangle,\\
    &&A_{3,ij}=\frac{\Gamma}{4}\sum_\lambda\rho_\lambda\left\langle
    \frac{\hat\gamma_i\hat\gamma_j(2\tilde\omega_n^2+|D_\lambda|^2)}{\Omega_n^3}\right\rangle,\quad
    A_{4,ij}=\frac{\Gamma}{4}\sum_\lambda\rho_\lambda\left\langle
    \frac{\hat\gamma_i\hat\gamma_jD_\lambda^2}{\Omega_n^3}\right\rangle,\\
    &&X_{0,i}=-\mu_B\sum_jA_{1,ij}H_j,\quad Y_{0,i}=-2\mu_B\sum_jA^*_{2,ij}H_j,
\end{eqnarray*}
and $\Omega_n=\sqrt{\tilde\omega_n^2+|D_\lambda|^2}$. Due to fast convergence, the Matsubara summation in Eq. (\ref{chi ij impurities gen}) can 
be extended to all frequencies.

\subsection{Residual susceptibility for isotropic pairing}
\label{sec: residual chi}

The general expression for the spin susceptibility, Eq. (\ref{chi ij impurities gen}), is rather cumbersome. On the other hand, application
of our results to real noncentrosymmetric materials is complicated by the lack of information about the superconducting
gap symmetry and the distribution of the pairing strength between
the bands. As an illustration, we focus on the isotropic pairing model introduced in Sect. \ref{sec: isotropic model}.
In this model, the order parameter magnitudes in the two bands are the same, but the relative phase can be either $0$ or $\pi$. 
While in the clean limit the spin susceptibility for both states is given by Eq. (\ref{chi ij SC final}), the
effects of impurities have to be analyzed separately.

\underline{$\eta_+=\eta_-=\eta$}.
Solving equations (\ref{XY eqs gen}) we obtain, in the coordinate system in which
$\langle\hat\gamma_i\hat\gamma_j\rangle$ is diagonal, the
following expression for the nonzero components of the
susceptibility tensor:
\begin{equation}
\label{chi-isotropic-singlet}
    \frac{\chi_{ii}(T)}{\chi_P}=1-\langle\hat\gamma_i^2\rangle
    \pi T\sum_n\frac{\eta^2}{\omega_n^2+\eta^2}\frac{1}{\sqrt{\omega_n^2+\eta^2}
    +\Gamma_i},
\end{equation}
where $\Gamma_i=(1-\langle\hat\gamma_i^2\rangle)\Gamma$.

\begin{figure}
    \includegraphics[width=9.0cm]{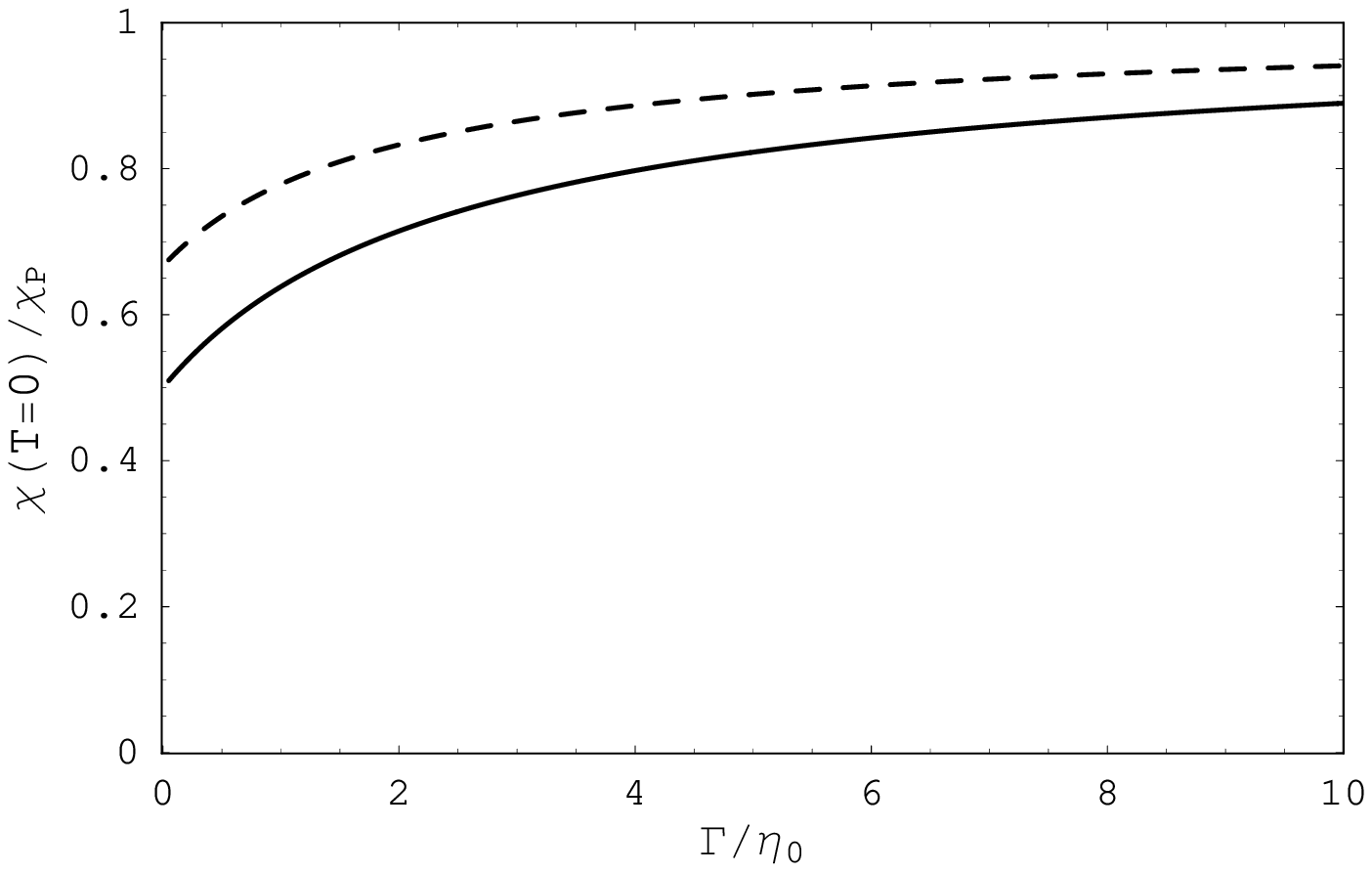}
    \caption{The residual susceptibility at $T=0$ vs disorder strength for $\eta_+=\eta_-$.
    The solid line corresponds to the transverse components in the
    2D case ($\chi_{zz}=\chi_P$ and is disorder-independent), the dashed line
    -- to the diagonal components in the 3D case.}
    \label{fig: chi_s}
\end{figure}

We are particularly interested in the effect of disorder on the
residual susceptibility at zero temperature. In this limit, the Matsubara sum
in Eq. (\ref{chi-isotropic-singlet}) can be replaced by a frequency integral, which
gives
\begin{equation}
\label{chi zero T singlet}
    \frac{\chi_{ii}(T=0)}{\chi_P}=
    1-\langle\hat\gamma_i^2\rangle+\langle\hat\gamma_i^2\rangle
    \Phi_1\left(\frac{\Gamma_i}{\eta_0}\right),
\end{equation}
where $\eta_0=(\pi/e^C)T_c$ is the gap magnitude at $T=0$, and
$$
    \Phi_1(x)=1-\frac{\pi}{2x}\left(1-\frac{4}{\pi\sqrt{1-x^2}}\arctan\sqrt{\frac{1-x}{1+x}}\right).
$$
While the first two terms on the right-hand side of Eq. (\ref{chi zero T singlet})
represent the residual susceptibility in the clean case, the last term describes the impurity effect.
In a weakly-disordered superconductor, using the asymptotics
$\Phi_1(x)\simeq\pi x/4$, we find that the residual susceptibility
increases linearly with disorder. In the dirty limit,
$\Gamma\gg\eta_0$, we have $\Phi_1(x)\to 1$, therefore
$\chi_{ii}(T=0)$ approaches the normal-state value $\chi_P$. For
the two simple band-structure models (2D and 3D) discussed earlier in this section, the Fermi-surface averages can be calculated
analytically, and we obtain the results plotted in Fig. \ref{fig: chi_s}.

Thus we see that, similarly to spin-orbit impurities in a usual
centrosymmetric superconductor,\cite{AG62} scalar impurities in a
noncentrosymmetric superconductor lead to an enhancement of the
spin susceptibility at $T=0$. Since the interband contribution is
not sensitive to disorder, this effect can be attributed to an
increase in the intraband susceptibilities.

\underline{$\eta_+=-\eta_-=\eta$}. From Eqs. (\ref{chi ij impurities gen}) and (\ref{XY eqs gen}) we obtain
the nonzero components of the susceptibility tensor:
\begin{equation}
\label{chi ii triplet}
    \frac{\chi_{ii}(T)}{\chi_P}=1-\langle\hat\gamma_i^2\rangle
    \pi T\sum_n\frac{\eta^2}{(\tilde\omega_n^2+\eta^2)^{3/2}
    -\Gamma\langle\hat\gamma_i^2\rangle\eta^2}.
\end{equation}
We note that for a spherical 3D model with $\langle\hat\gamma_i^2\rangle=1/3$ this expression has exactly
the same form as the susceptibility of the superfluid ${}^3$He-B
in aerogel, see Refs. \cite{MK01} and \cite{SS01}.

\begin{figure}
    \includegraphics[width=9.0cm]{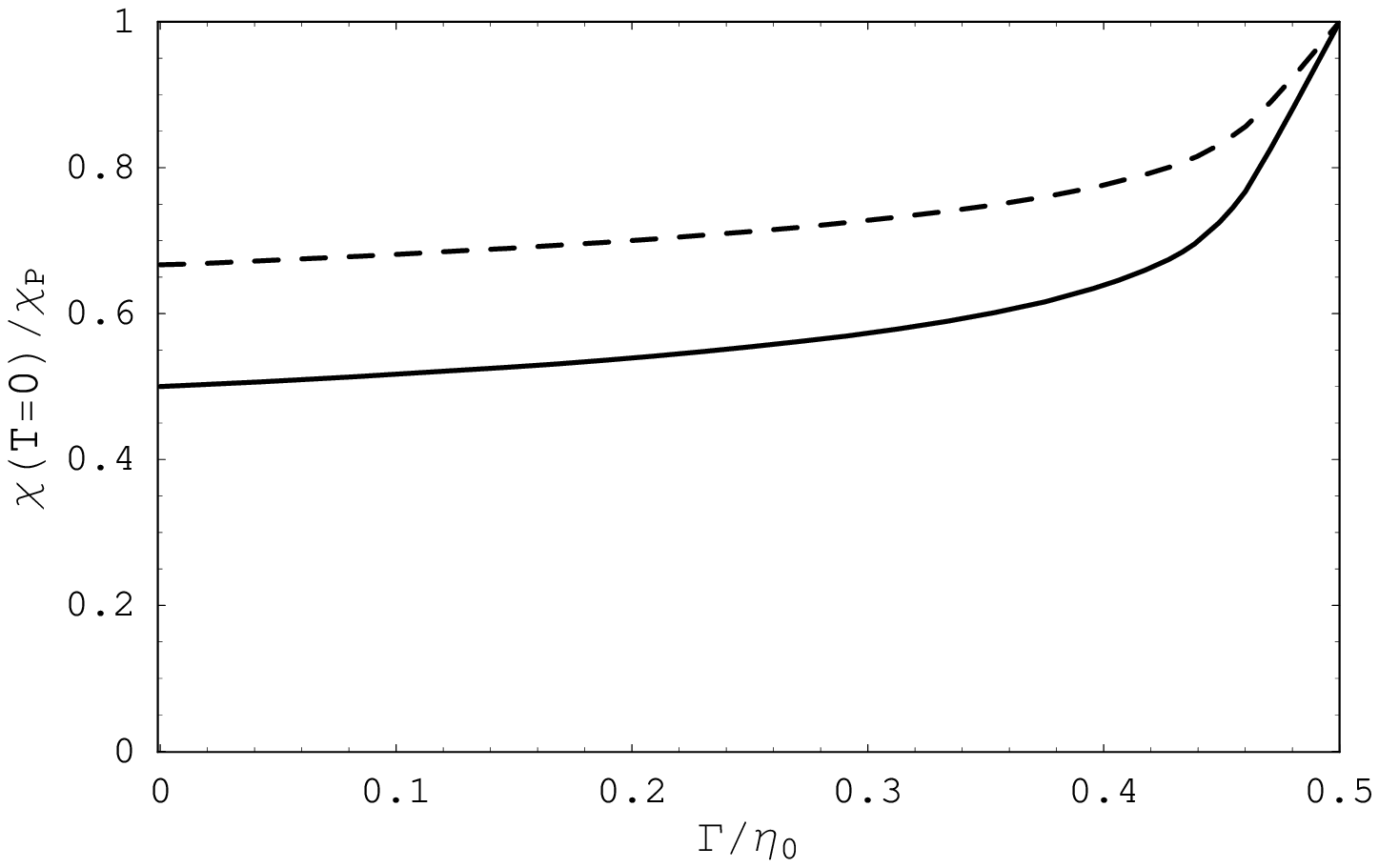}
    \caption{The residual susceptibility at $T=0$ vs disorder strength for $\eta_+=-\eta_-$.
    The solid line corresponds to the transverse components in the
    2D case ($\chi_{zz}=\chi_P$ and is disorder-independent), the dashed line
    -- to the diagonal components in the 3D case.}
    \label{fig: chi_t}
\end{figure}

At $T=0$, the expression (\ref{chi ii triplet}) takes the following form:
\begin{equation}
\label{chi zero T triplet}
    \frac{\chi_{ii}(T=0)}{\chi_P}=1-\langle\hat\gamma_i^2\rangle
    +\langle\hat\gamma_i^2\rangle\Phi_2\left(\frac{\Gamma}{\eta}\right),
\end{equation}
where
$$
    \Phi_2(x)=1-\int_{y_{min}}^\infty dy
	\left[1-\frac{x}{(y^2+1)^{3/2}}\right]\frac{1}{(y^2+1)^{3/2}-x\langle\hat\gamma_i^2\rangle},
$$
and $y_{min}=\theta(x-1)\sqrt{x^2-1}$. The last term on the
right-hand side of Eq. (\ref{chi zero T triplet}) describes the
effect of impurities. According to Sect. \ref{sec: isotropic model}, superconductivity is suppressed above
the critical disorder strength $\Gamma_c=(\pi/2e^C)T_{c0}$.
For a given $\Gamma$, one should first obtain the gap magnitude from Eq. (\ref{eq for eta}) and then calculate $\Phi_2(x)$. In the
weak disorder limit, we have $\Phi_2(x)\simeq(3\pi x/16)(1-\langle\hat\gamma_i^2\rangle)$, i.e. the residual
susceptibility increases linearly with disorder. At $\Gamma\to\Gamma_c$, we have $\Phi_2(x)\to 1$ and $\chi_{ii}(T=0)\to\chi_P$.
The dependence of $\chi_{ii}(T=0)$ on the disorder strength for
the 2D and 3D models is plotted in Fig. \ref{fig: chi_t}. As in
the case $\eta_+=\eta_-$, the residual susceptibility is enhanced
by impurities.

\section{Conclusions}
\label{sec: conclusions}

Scalar disorder in noncentrosymmetric superconductors causes anisotropic mixing of the electron states in the bands split by the SO coupling. 
The critical temperature is generally suppressed by impurities, but this happens differently for conventional and unconventional pairing. For all types of unconventional pairing (which is defined as corresponding to a non-unity representation of the crystal point group, with vanishing Fermi-surface
averages of the gap functions), the impurity effect on $T_c$ is described by the universal Abrikosov-Gor'kov equation. The same is also true for certain types
of conventional pairing, in particular the ``protected'' isotropic triplet state with $\eta_+=-\eta_-$. Any deviation from the Abrikosov-Gor'kov curve, 
in particular, an incomplete suppression of superconductivity by strong disorder, is a signature of conventional pairing symmetry. 

The impurity-induced mixing of singlet and triplet pairing channels makes the magnetic response of noncentrosymmetric superconductors
with the SO coupling different from the centrosymmetric case. In an isotropic BCS-like model, the upper critical field $H_{c2}$ is enhanced by disorder at all temperatures, the magnitude of the effect depending on the SO coupling strength. In general, the effect of impurities on the slope of $H_{c2}$ is sensitive to the pairing symmetry and the band structure. 

Concerning the spin susceptibility, we found that scalar impurities in noncentrosymmetric superconductors act similarly to spin-orbit impurities in centrosymmetric
superconductors, in the sense that they enhance the residual susceptibility at $T=0$. The quantitative details again
depend on the band structure, the anisotropy of the SO coupling, and the symmetry of the order parameter.

Most of the experimental work on noncentrosymmetric superconductors has been done on CePt$_3$Si. In this compound 
the Fermi surface is quite complicated and consists of multiple sheets.\cite{SZB04} It is not known which of them are superconducting. 
The order parameter symmetry is not known either, although there is experimental evidence that there are lines of nodes in the gap.\cite{Yasuda04,Izawa05,Bon09}
The data on the impurity effects are controversial. The experimental samples seem to be rather clean, with the ratio of the elastic mean
free path $l$ to the coherence length $\xi_0$ ranging from $4$ (Ref. \cite{Bauer04}) to $10-27$ (Ref. \cite{Yogi06}. There are indications that $T_c$ is indeed suppressed by structural defects and/or impurities in some samples.\cite{Bon09} On the other hand, the values of both the critical temperature and the
upper critical field in polycrystalline samples\cite{Bauer04} are higher than in single crystals.\cite{Yasuda04} This is opposite to what has been observed in 
other unconventional superconductors and also disagrees with the theoretical predictions, assuming that the polycrystals
are intrinsically more disordered than the single crystals. In addition, the low-temperature behaviour of the penetration depth in disordered samples is unusual\cite{Bon09} and cannot be explained by existing theoretical models.  
In order to resolve these issues, more systematic studies of the disorder effects in a wide range of impurity concentrations are needed.

\section*{Acknowledgments}

The author is grateful to D. Agterberg, I. Bonalde, S. Fujimoto, V. Mineev, B. Mitrovic, and M. Sigrist for useful discussions.
This work was supported by a Discovery Grant from the Natural Sciences and Engineering Research Council of Canada.

\end{document}